# Electronic Structure, Surface Doping and Optical Response in Epitaxial WSe$_2$ Thin Films

Yi Zhang[1, 2, 3, *], Miguel M. Ugeda[4,5,6], Chenhao Jin[4], Su-Fei Shi[4,7], Aaron J. Bradley[4], Ana Martín-Recio [4, 8], Hyejin Ryu[3, 9], Jonghwan Kim[4], Shujie Tang[10,11], Yeongkwan Kim[3], Bo Zhou[3, 10,12], Choongyu Hwang[9, 13], Yulin Chen[12], Feng Wang[4, 14, 15], Michael F. Crommie[4, 14, 15], Zahid Hussain[3], Zhi-Xun Shen[2, 10], Sung-Kwan Mo[3, *]

[1]*National Laboratory of Solid State Microstructures, School of Physics, Collaborative Innovation Center of Advanced Microstructures, Nanjing university, Nanjing 210093, China*
[2]*Stanford Institute of Materials and Energy Sciences, SLAC National Accelerator Laboratory, Menlo Park, CA 94025, USA*
[3]*Advanced Light Source, Lawrence Berkeley National Laboratory, Berkeley, CA 94720, USA*
[4]*Department of Physics, University of California at Berkeley, Berkeley, CA 94720, USA*
[5]*CIC nanoGUNE, 20018 Donostia-San Sebastian, Spain*
[6]*Ikerbasque, Basque Foundation for Science, 48011 Bilbao, Spain*
[7]*Department of Chemical and Biological Engineering, Rensselaer Polytechnic Institute, Troy, NY 12180, USA*
[8]*Departamento de Física de la Materia Condensada, Universidad Autónoma de Madrid, E-28049 Madrid, Spain*
[9]*Max Plank POSTECH Center for Complex Phase Materials, Pohang University of Science and Technology, Pohang 790-784, Korea*
[10]*Geballe Laboratory for Advanced Materials, Departments of Physics and Applied Physics, Stanford University, Stanford, CA 94305, USA*
[11]*State Key Laboratory of Functional Materials for Informatics, Shanghai Institute of Microsystem and Information Technology, Chinese Academy of Sciences, Shanghai 200050, China*
[12]*Department of Physics and Clarendon Laboratory, University of Oxford, Parks Road, Oxford, OX1 3PU, UK*
[13]*Department of Physics, Pusan National University, Busan 609-735, Korea*
[14]*Materials Sciences Division, Lawrence Berkeley National Laboratory, Berkeley, CA 94720, USA*
[15]*Kavli Energy NanoScience Institute at the University of California Berkeley and the Lawrence Berkeley National Laboratory, Berkeley, CA 94720, USA*
*Email: zhangyi@nju.edu.cn
          skmo@lbl.gov

**High quality WSe$_2$ films have been grown on bilayer graphene (BLG) with layer-by-layer control of thickness using molecular beam epitaxy (MBE). The combination of angle-resolved photoemission (ARPES), scanning tunneling microscopy/spectroscopy (STM/STS), and optical absorption measurements reveal the atomic and electronic structures evolution and optical response of WSe$_2$/BLG. We observe that a bilayer of WSe$_2$ is a direct bandgap semiconductor, when integrated in a BLG-based heterostructure, thus shifting the direct-indirect band gap crossover to trilayer WSe$_2$. In the monolayer limit, WSe$_2$ shows a spin-splitting of 475 meV in the valence band at the K point, the largest value observed among all the MX$_2$ (M = Mo, W; X = S, Se) materials. The exciton binding energy of monolayer-WSe$_2$/BLG is found to be 0.21 eV, a value that is orders of magnitude larger than that of conventional 3D semiconductors, yet small as compared to other 2D transition metal dichalcogennides (TMDCs) semiconductors. Finally, our finding regarding the overall modification of the electronic structure by an alkali metal surface electron doping opens a route to further control the electronic properties of TMDCs.**

Two-dimensional (2D) transition metal dichalcogenides (TMDCs) $MX_2$ (M = Mo, W; X = S, Se) semiconductors have attracted extensive interest due to their remarkable fundamental properties distinct from those of their bulk counterparts.[1-4] These include direct electronic bandgap in the single-layer limit,[5-10] spin-splitting of the valence band (VB),[11-15] a well-pronounced valley degree of freedom[16-20] as well as the excitonic nature of their optical spectra.[20-24] Large efforts are currently devoted to tailor these properties in order to widen further the versatility of these materials and to achieve practical devices out of them.[25-38] Some proposed strategies in these regards are the growth of films with varying thickness,[14, 39-42] their growth on other 2D materials to form hybrid heterostructures,[28, 31, 35, 37, 43] and the use of chemical and electrostatic doping.[13, 29, 44-46] All of them have been demonstrated to be very successful in modifying and controlling the properties of graphene, which have even led to the observation of novel physical phenomena.[47-51] Nonetheless, these strategies have been less explored in TMDC materials although promising results have recently been predicted.[37]

Among the family of TMDC semiconductors, $WSe_2$ is probably the most interesting material for potential applications. It is expected to possess the largest spin-splitting in the VB at the K/K' point among all the $MX_2$ semiconductors,[11] which makes $WSe_2$ an ideal platform for studying spin and valley dependent properties as well as for spintronic applications.[18, 44] The recent polarization dependent photoluminescence (PL) indicates that the valley coherence is preserved for longer time compared to other TMDCs,[52] which makes $WSe_2$ a more promising candidate for valleytronics. So far, detailed spectroscopic research on the electronic structure of monolayer $WSe_2$ is rather scarce due to the difficulty in obtaining high-quality thin films and hybrid heterostructures with uniform thickness,[53] which, to date, can be only obtained via chemical vapor deposition[42] and, in a lesser extent, via molecular beam epitaxy (MBE).[8]

Here we report the MBE growth and subsequent characterization of hybrid heterostructures formed by high-quality one-atomic-plane precision films of $WSe_2$, with varying thicknesses from one to eight monolayers (MLs), on a bilayer graphene (BLG) substrate. Combining *in-situ* angle-resolved photoemission spectroscopy (ARPES), optical absorption, scanning tunneling microscopy/spectroscopy (STM/STS), core level spectroscopy, low energy electron diffraction (LEED) and reflection high-energy electron diffraction (RHEED) techniques, we study the atomic and electronic structures evolution and optical response of these heterostructures. Remarkably, we find that a bilayer of $WSe_2$ remains a direct bandgap semiconductor, when is part of a BLG-based heterostructure, thus shifting the direct-indirect bandgap crossover to the trilayer of $WSe_2$. Furthermore, our ARPES spectra show a rather large spin-splitting of 475 meV in the VB at the K points of the Brillouin zone of the 1-ML-$WSe_2$/BLG heterostructure. We also present unambiguous experimental measurement of the binding energy for neutral excitons in this heterostructure. We obtain an exciton binding energy of 0.21 eV for this TMDC semiconductors on BLG, a value that, despite is orders of magnitude larger than that observed in conventional 3D semiconductors, is yet intriguingly smaller as compared to other TMDCs.[22-24] Lastly, we analyze the evolution of the size and the character of the electronic band gap with chemical surface doping. Overall, our results provide a well-defined route to create high-quality large-scale $WSe_2$/BLG heterostructures as well as new avenues to tailor the electronic and optoelectronic properties of TMDCs.

Figures 1a-c show the crystal structure of $WSe_2$. A single layer (Se-W-Se) of $WSe_2$ consists of two planes of Se atoms separated by one layer of W atoms in a trigonal prismatic coordination. Layers of $WSe_2$ are vertically stacked by van der Waals interactions in an AB configuration. Figure 1d shows the 2D Brillouin zone of the $WSe_2$ layers. Since the bandgap of $MX_2$ semiconductors is along the Γ-K direction, we will focus on this direction in the following ARPES measurements. BLG is an ideal substrate for epitaxial growth of layered materials such as $Bi_2Se_3$ and $MoSe_2$ due to its honeycomb atomic structure and van der Waals nature.[8, 40, 54] An

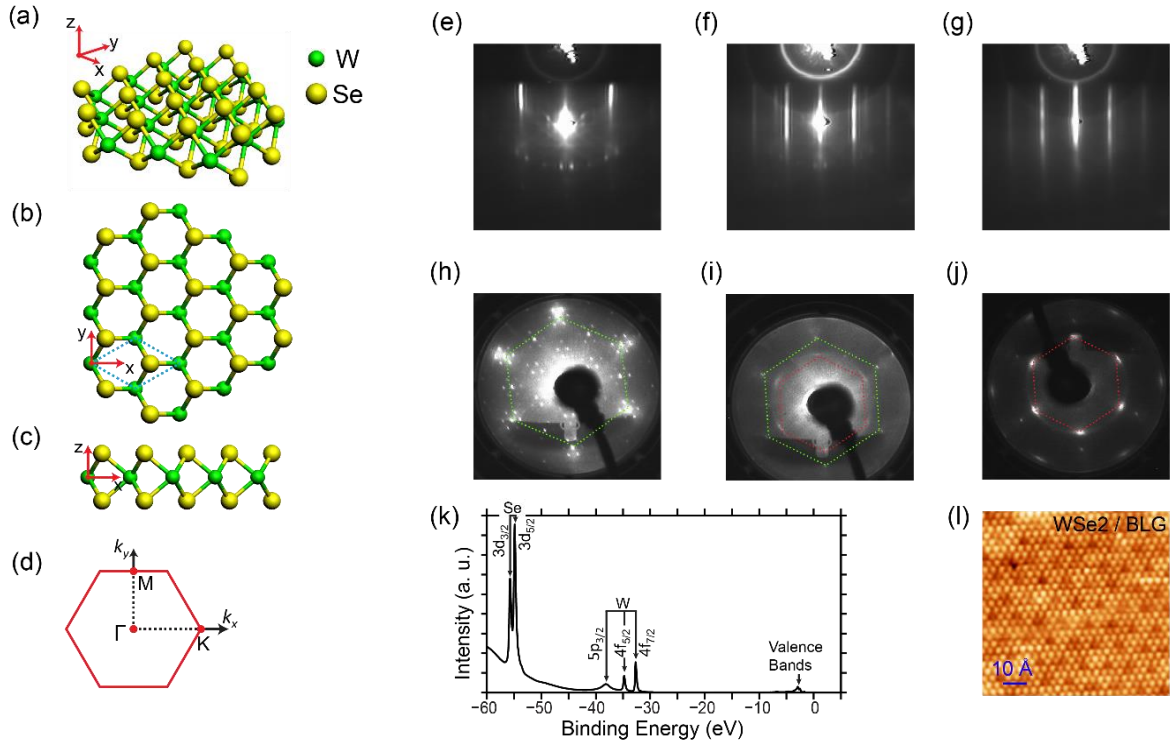

**Figure 1 | Growth of WSe$_2$ thin films. (a)-(c)** Crystal structure of 2D WSe$_2$ with **(a)** perspective view, **(b)** top view, and **(c)** side view, respectively. **(d)** 2D Brillouin zone of WSe$_2$ film. **(e)-(g)** RHEED pattern along the Γ-K direction of **(e)** BLG substrate, **(f)** 0.7 ML WSe$_2$ film, and **(g)** monolayer WSe$_2$ film, respectively. **(h)-(j)** LEED pattern of **(h)** BLG substrate, **(i)** 0.7 ML WSe$_2$ film, and **(j)** monolayer WSe$_2$ film, respectively. The dotted green and red hexagon indicates 1×1 diffraction pattern of BLG substrate and epitaxial WSe$_2$ film, respectively. **(k)** Core-level spectrum of epitaxial WSe$_2$ film. **(i)** STM image of WSe$_2$ film (V$_s$ = - 1.3 V, I$_t$ = 0.6 nA, T = 5 K).

important advantage of BLG substrate for high-quality growth is that the lattice constant ratio between BLG and WSe$_2$ is very close to 3:4,[7] which facilitates WSe$_2$ to form a single-crystalline thin film. To prepare an uniform BLG substrate, a 6H-SiC(0001) wafer was first initially degassed at 650 °C in the ultra-high vacuum (UHV) chamber for 3 hours, followed by 80 cycles of flash-annealing to 1300 °C.[55] Figures 1e and h show the RHEED and LEED patterns of BLG substrate, respectively. The characteristic sharp diffraction patterns of graphene indicate the high quality of the substrate. For the growth of WSe$_2$ thin film, high-purity W and Se were evaporated from an electron-beam evaporator and a standard Knudsen cell, respectively. The flux ratio of W:Se was controlled to be ~ 1:30. Excess amount of Se was deposited to avoid Se vacancy and W cluster nucleation in the film. The BLG substrate was kept at ~ 400 °C during the growth. This substrate temperature leads a stoichiometric crystallization of WSe$_2$ film. After growth, we post-annealed the film at ~ 550 °C under Se flux for 10 minutes to improve the crystalline quality. The film thickness was accurately controlled by the deposition time, with a growth rate of 17 minutes per monolayer, monitored by the *in-situ* RHEED. Then, afterwards, its crystal orientation and quality was also checked by *in-situ* LEED.

Figures 1f and i show the RHEED and LEED patterns of the WSe$_2$ film for a coverage of 0.7 ML (0.7 ML means that a 70% area of the substrate surface was covered by monolayer WSe$_2$), respectively. In these patterns, both the BLG and WSe$_2$ diffraction spots can be observed. The LEED pattern shows the diffraction spots of WSe$_2$ aligned with those of the BLG substrate although slightly stretched along the rotational direction. This reveals that the first layer of WSe$_2$ has the same atomic lattice orientation as the BLG substrate although it also presents domains with small rotational misorientation (θ < ± 4°). When the coverage reached 1 ML, the RHEED pattern of BLG vanishes (Figure 1g), and only the WSe$_2$ pattern can be observed. In the corresponding LEED pattern (Figure 1j), the pattern of BLG is barely visible and much weaker than that of the WSe$_2$.

We also performed core level spectroscopy and low temperature (5K) STM measurements for a thorough characterization of the atomic structure of the WSe$_2$ films. Figure 1k shows a typical angle-integrated core level spectrum taken by the *in-situ* ARPES system. The sharp characteristic peaks of Se (54.7 eV of 3d5/2 orbit and 55.5 eV of 3d3/2 orbit) and W (32.5 eV of 4f7/2 orbit, 34.6 eV of 4f5/2 orbit and 37.9 eV of 5p3/2 orbit) indicate the 2:1 stoichiometry and demonstrate the purity of the WSe$_2$ films. The atomically resolved STM image of monolayer WSe$_2$ on BLG (Figure 1l) shows simultaneously an atomic periodicity of 3.29 Å and a moiré pattern formed between the graphene and the WSe$_2$ atomic lattices. Similar to the epitaxial MoSe$_2$ on BLG/SiC(0001), the moiré pattern in 1-ML-WSe$_2$/BLG is a (3×3) structure with respect to WSe$_2$ with a periodicity between 9.9 Å to 9.3 Å,[23] which confirms the aligned orientation of the WSe$_2$/BLG heterostructure deduced from LEED patterns (Figures 1i and j).

For a systematic characterization of the electronic properties of variable-thickness WSe$_2$/BLG heterostructures, we first explored their band structure by performing *in-situ* ARPES measurements. Figures 2a-d present the ARPES spectra of the epitaxial WSe$_2$ thin films along the Γ-K direction, with varying film thicknesses of 1 ML, 2 ML, 3 ML and 8 ML. Figure 2e-h are the corresponding second-derivatives of the original spectra in Figures 2a-d for enhanced visibility of the band structures. The contribution from BLG is out of this momentum window (*Supporting Information A*). The top of the VBs, depicted by red dashed lines in Figures 2a-d, show different number of the branches at the Γ point following the number of layers: for monolayer WSe$_2$ the Γ point shows only one branch at the top of the VB. It splits into two branches for 2 ML, and then into three branches in the 3 ML film. For the 8 ML WSe$_2$ film, theory suggests 8 branches to appear,[7] but we can only observe broadened multiple bands due to the limited resolution and narrower spacing of branches in energy. Since the number of branches in the VB near the Γ point corresponds to the number of layers, this can be used as unique identifier of the thickness of the ultra-thin WSe$_2$ films.

To gain further information of the VB evolution with varying film thickness, we present zoom-in ARPES spectra focusing only on the top of the VB in Figures 2i-l. The most important feature in the evolution of the band structure of MX$_2$ semiconductors is the indirect to direct bandgap transition. This bandgap transition is concomitant with the change of the valence band maximum (VBM) from the Γ point to the K point in the monolayer limit as predicted in theoretical calculations[7] and observed in previous ARPES measurements for MoSe$_2$ and MoS$_2$.[8,9] Our ARPES spectra on monolayer WSe$_2$ show that the VBM is located at the K point and 0.56 ± 0.01 eV higher than the top of the VB at the Γ point. The precise energy positions of each band were assigned from a fitting using multiple Gaussian peaks for the energy distribution curves (EDCs) at the Γ and K points (*Supporting Information B*). The energy difference between the K and Γ point in the VB ($E_K$-$E_\Gamma$ = 0.56 eV) for monolayer WSe$_2$ is significantly larger than that of monolayer MoSe$_2$ (0.38 eV)[8] and MoS$_2$ (0.31 eV)[14]. The larger difference ($E_K$-$E_\Gamma$) implies the strongest tendency of monolayer WSe$_2$ to maintain a direct bandgap among all the MX$_2$ (M = Mo, W and X = S, Se) semiconductors.[7] The transition to an indirect bandgap has clearly been observed in 2 ML MoSe$_2$ and MoS$_2$.[8,9] In contrast, our ARPES spectra show that the top of VB at the K point is still 0.08 eV higher than that at the Γ point for 2 ML WSe$_2$ on BLG, thus enabling the possibility that the 2 ML WSe$_2$ in the heterostructure may be a direct bandgap semiconductor. For 3 ML and 8 ML WSe$_2$, the top of the VB at the Γ point is slightly higher than that at the K point and becomes the VBM, which suggests that the direct to indirect bandgap transition may occur between 2 ML and 3 ML in WSe$_2$.

Another key property of MX$_2$ semiconductors is the spin-split band structure due to strong spin-orbit coupling and inversion symmetry breaking.[11] Monolayer WSe$_2$ has been theoretically predicted to have the largest splitting size among all the MX$_2$ semiconductors.[11] As shown in the ARPES spectra of Figure 2i, the VB at the K point splits into two branches separated by 475 ± 5 meV, which is much larger than that measured for monolayer MoSe$_2$ (~180 meV, ref. 8) and MoS$_2$ (~150 meV, ref. 14). This is due to the largely enhanced spin-orbit coupling in W as compared to that in Mo and makes WSe$_2$ particularly promising among all the TMDC compounds for spintronic applications.

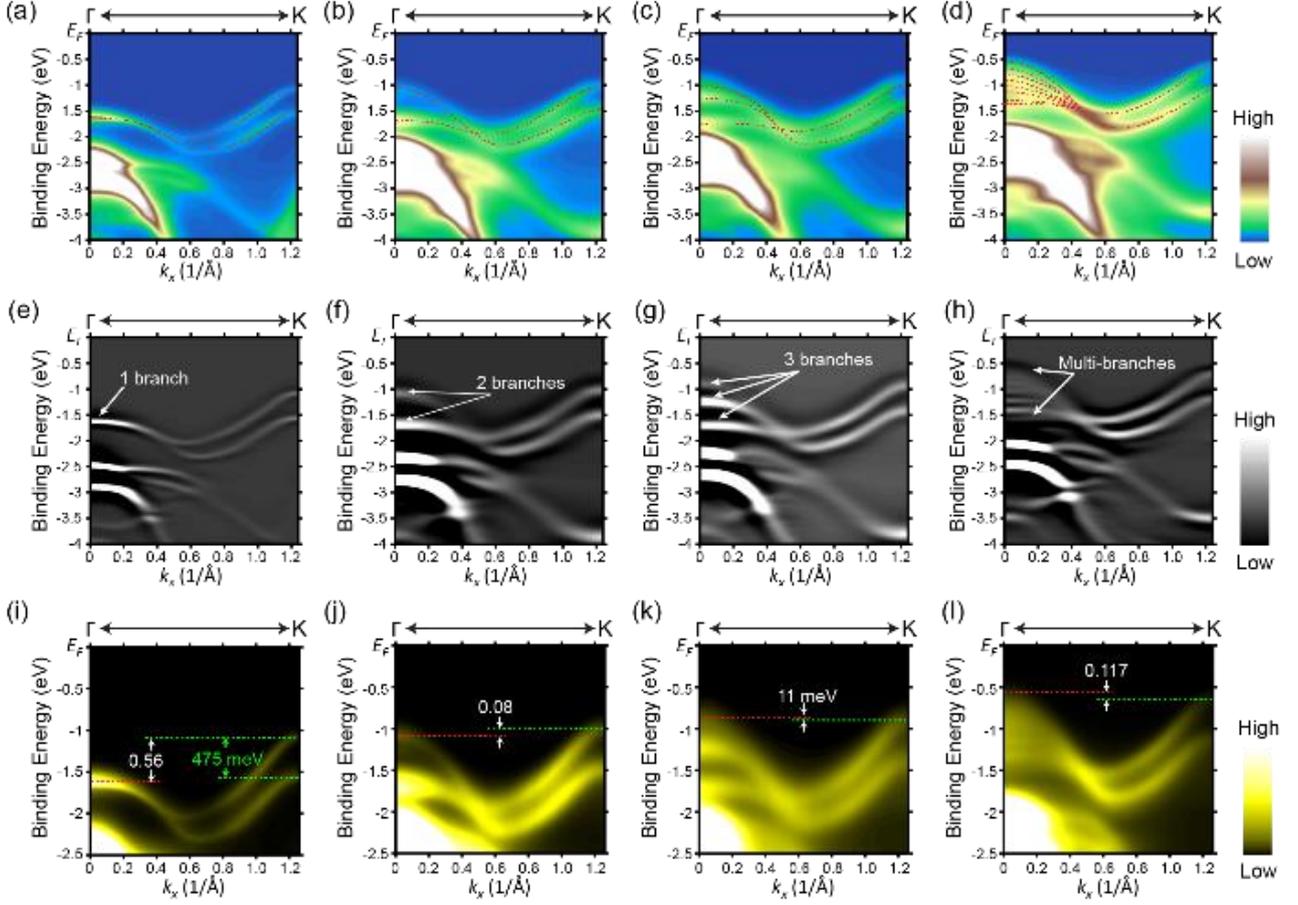

**Figure 2 | ARPES data of epitaxial WSe$_2$ thin films. (a)-(d)** ARPES spectra of **(a)** 1 ML, **(b)** 2 ML, **(c)** 3 ML and **(d)** 8 ML WSe$_2$ films along the Γ-K direction. The red dotted curves depict the top VBs. **(e)-(h)** Second-derivative ARPES spectra of **(e)** 1 ML, **(f)** 2 ML, **(g)** 3 ML and **(h)** 8 ML WSe$_2$ films along the Γ-K direction. **(i)-(j)** Zoom-in ARPES spectra of **(i)** 1 ML, **(j)** 2 ML, **(k)** 3 ML and **(l)** 8 ML WSe$_2$ films along the Γ-K direction. The red and green dotted lines indicate the energy positions of top of VB at the Γ and K point, respectively. All the labeled numbers have unit of eV.

To evaluate the bandgap character – direct or indirect - of the WSe$_2$ films on BLG, we study the conduction band minimum (CBM) location in energy and momentum by controlled surface doping of the WSe$_2$ with alkali metal (Na and K) at 60K (*Supporting Information C*). This procedure allows us to shift the Fermi level (E$_F$) upwards to make the CBM accessible for ARPES.[56] Although we have used both Na and K adatoms to this purpose, only K doping enabled us to shift the CBM below E$_F$. ARPES spectra of K-doped monolayer WSe$_2$ on BLG (Figure 3a) reveals that the CBM becomes visible at the K point, as expected for a single layer of a MX$_2$ semiconductors, and the VB shifts downwards 0.56 eV and 0.47 eV at the Γ and K point, respectively. Therefore, the monolayer K-doped WSe$_2$ shows a direct electronic bandgap of 1.40 ± 0.02 eV. In the ARPES spectrum for K-doped 2 ML-WSe$_2$ (Figure 3b), while the CBM remains at the K point, the VBM now shifts to the Γ point, which leads to an indirect bandgap of 1.26 eV. Similar energetics have been observed for the K-doped 8 ML WSe$_2$ (Figure 3c), which shows an indirect bandgap of 1.01 eV. These gap values are much smaller than that expected for single- and few-layer MX$_2$ semiconductors and even for MX$_2$/BLG heterostructures (ref. 23 and 57). This is caused by a doping-induced enhancement of charge screening at the semiconductor, which leads to the band structure renormalization.[45]

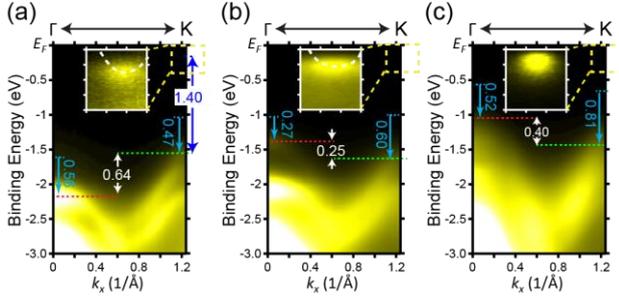

**Figure 3 | Surface doping effect of WSe$_2$ films. (a)-(c)** ARPES spectra of **(a)** 1 ML, **(b)** 2 ML and **(c)** 8 ML WSe$_2$ films with K surface doping. The red and green dotted lines indicate the energy positions of top of VB at the Γ and K point, respectively. Cyan arrows and numbers show the band movements at the Γ and K point after surface doping comparing to undoped films. Insets in **(a), (b) and (c)** are the zoom-in spectra with 10 times enhanced intensity to make the CBM visible. All the labeled numbers have unit of eV.

In contrast with our experimental results, previous calculations have shown that the CBM of isolated WSe$_2$ from 2 ML to bulk is located at the mid-point between the Γ and K point (Q point).[7, 10] This discrepancy can be attributed to the strain effect induced by the BLG substrate due to the lattice mismatch between the film and the substrate.[58] This affects the momentum location of CBM, the energy difference between the K and Γ point in the VB ($E_K$-$E_Γ$), as well as the splitting at the K point.[26, 32, 33] Furthermore, the difference in energy of the VB between the K and Γ points ($E_K$-$E_Γ$ = 0.56 eV) and the energy splitting at the K point (475 meV) in monolayer WSe$_2$ are both smaller than those found for the exfoliated WSe$_2$ (0.89 eV and 513 meV),[53] which is also suggestive of strain induced in WSe$_2$ by the BLG in the heterostructure.

Our ARPES measurements on K-doped WSe$_2$/BLG reveal that the CBM remains at the K point from 1 ML up to 8 ML of WSe$_2$. Combining this observation with our ARPES spectra on VBs of pristine WSe$_2$ films, we suggest that the direct-indirect bandgap transition occurs from 2 ML to 3 ML for undoped WSe$_2$ epitaxial films on BLG. Since the energy difference between the VBs at Γ and K points is small ($E_Γ$-$E_K$ ~0.01 eV) for 2 ML and 3 ML WSe$_2$, and the in-plane strain could also renormalize the VB of MX$_2$ semiconductors, the crossover thickness of the direct-indirect bandgap transition could be engineered by strain. For the surface doping effect, the momentum position inversion in VBM for K-surface-doped 2 ML WSe$_2$ implies a new method to control the direct-indirect bandgap transition via surface chemical doping. These observations paint a clear picture of the effect of chemical doping and strain on the electronic structure of thin films of MX$_2$ semiconductors integrated in hybrid heterostructures.

Due to the changes introduced in the band structure in surface-doped thin films, ARPES is not an accurate tool to measure the quasiparticle bandgap of pristine TMDCs. In order to investigate the fundamental optical and electronic bandgaps of pristine WSe$_2$, we performed both optical absorbance and high-resolution STS on our epitaxial monolayer WSe$_2$ thin films. This combined experimental approach also allows us to obtain an accurate and unambiguous value for the exciton binding energy, a critical parameter for understanding how light interacts with TMDC materials, in particular regarding processes such as its optical absorption and photovoltaic response.

Figure 4a shows the optical absorbance spectrum of the monolayer WSe$_2$ film taken at 77 K. The A exciton peak is broad, likely due to charge transfer and energy transfer between the WSe$_2$ and the graphene, but clearly resolved with the center at 1.74 ± 0.01 eV (713 nm). This is consistent with previous reports on exfoliated WSe$_2$ monolayers.[59] The absorption signal around the B exciton peak is even broader and centered around 2.17 eV (~ 570 nm). The energy difference between the two absorption peaks (~ 0.43 eV) agrees with the band splitting energy observed in the ARPES spectra in Figure 2i. The absorption spectrum at room temperature is also provided in *Supporting Information D*. Measurement of the electronic bandgap (E$_g$) of undoped single-layer WSe$_2$ was performed by STS at T = 5 K. Typical STS *dI/dV* spectrum acquired on monolayer WSe$_2$ exhibits large electronic bandgap (E$_g$) around E$_F$, as shown in Figure 4b. The locations of the VBM and the CBM and, therefore, the value of E$_g$ were obtained by means of statistical analysis using a sample of N = 30 dI/dV curves and following the fitting procedure detailed in Ref. 23. VBM in monolayer WSe$_2$ is found to be located at -1.10 ± 0.02 V, which nicely agrees with our ARPES result, and the CBM at +0.85 ± 0.03 V. The asymmetry of the VBM and CBM respect to E$_F$ reveals a slight n-type doping of our WSe$_2$ films, albeit with a negligible carrier concentration. The nearly intrinsic character of our epitaxial WSe$_2$ films suggests a high crystal quality, in accordance with the low concentration of point defects found from our STM images (*Supporting Information E*), which are known to be a source of doping in TMDC materials.[30] Our statistical analysis of the STS spectra yields a value for the single-particle electronic bandgap of E$_g$ = E$_{CBM}$ – E$_{VBM}$ = 1.95 ± 0.04 eV. For a direct bandgap semiconductor such as monolayer WSe$_2$, the difference between the electronic bandgap and the optical bandgap represents the exciton binding energy (E$_b$), which in the present case we have found to be E$_b$ = E$_g$-E$_{opt}$ = 0.21 ± 0.04 eV.

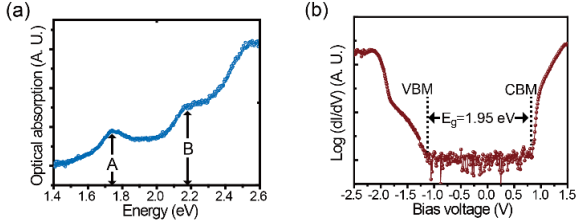

**Figure 4 | Optical and electronic bandgap of epitaxial monolayer WSe$_2$. (a)** Optical absorption spectrum taken on monolayer WSe$_2$ film. The two main absorption peaks A and B are indicated. **(b)** STS spectrum taken on monolayer WSe$_2$ film (f = 871 Hz, I$_t$ = 50 nA, V$_{rms}$ = 1.5 mV, T = 5 K).

This large exciton binding energy is explained by enhanced Coulomb interactions in low dimensional systems.[20-23] The large dielectric constant in bulk semiconductors gives rise to reduced strength of the Coulomb interaction and, therefore, limits the binding energies of these excitations within few meV. In 2D systems, reduced screening leads to enhanced Coulomb interactions, which significantly increase the binding energy of electron-hole excitations.[22] We have recently demonstrated these effects for 2D TMDC and reported an exciton binding energy of 0.55 eV for single layer of epitaxial MoSe$_2$ film on BLG.[23] Using this method, we have been able to extract a large exciton binding energy of 0.21 eV for epitaxially grown monolayer WSe$_2$ on BLG, that is, however, significantly smaller than that of monolayer MoSe$_2$ film.[23] This value also results smaller than the exciton binding energy of single-layer WSe$_2$ on insulating environments such as SiO$_2$ estimated by optical methods, which ranges from 0.37 eV[59] up to 0.6 eV[60], likely due to the increased screening substrate (BLG) environment.

This exciton binding energy is several times larger than those extracted for bulk TMDCs[22], and, therefore, demonstrates the enhanced coulomb interactions in the monolayer limit for WSe$_2$. However, this value is 62% smaller than that measured for monolayer MoSe$_2$ on the same substrate.[23] The smaller exciton binding energy is caused by both a larger electronic gap and smaller optical bandgap than those found in MoSe$_2$. This difference is, therefore, not caused by the dielectric environment but has its origin in the distinct metal atom between the two compounds. It is worth noting that a similar binding energy has been reported on WSe$_2$ on SiO$_2$ substrate[59]. This drastic difference in the exciton binding energy among the different TMDCs is very intriguing itself and calls for further theoretical attention.

Interestingly, the electronic bandgap measured in ARPES (1.40 ± 0.02 eV) is significantly smaller than that measured by STS (1.95 ± 0.04 eV). The different values of the bandgap measured by ARPES and STS are due to the surface doping. While the STS bandgap was measured on un-doped thin films, the bandgap measurement by ARPES was performed on the K-doped films. We suggest the K surface doping induces a bandgap renormalization due to extra free carriers on the film, which effectively increases charge screening in the material. This naturally reduces both the quasiparticle bandgap and the exciton binding energy. As it has been previously shown in Na-doped $MoS_2$, alkali metal surface doping on $MX_2$ reduces the bandgap by mainly affecting the CB rather than the VB.[45]

In conclusion, we successfully synthesized ultra-thin $WSe_2$ film on epitaxial BLG substrate with controllable thickness at the atomic level. *In-situ* ARPES measurements directly demonstrate the layer-by-layer electronic structure evolution of the epitaxial $WSe_2$ film, suggesting a direct bandgap in monolayer and 2 ML $WSe_2$, and indirect bandgap for 3 or more ML $WSe_2$ on BLG substrate. In monolayer $WSe_2$ film, we also observed a giant VB splitting (475 ± 5 meV) at the K point, which is larger than any other monolayer of $MX_2$. The further surface doping experiments show that the electronic structure undergoes a significant change, allowing us a further control of band structure of $WSe_2$, such as the size of the gap and direct-indirect bandgap transition. The exciton binding energy observed in monolayer $WSe_2$ highlights the importance of many-body effects in atomically-thin 2D layers and has a profound impact on future technologies involving single-layer semiconducting TMDCs, such as solar cells and valleytronic devices, either in stand-alone devices or within integrated heterostructures. Our MBE growth of $WSe_2$ and studies on the VB evolution, VB splitting at K point, surface doping effect, bandgap and excitonic effect not only help understanding of TMDC materials, but also enrich the family of epitaxial 2D materials towards a fully MBE grown epitaxial heterostructures for light emission and photon-voltage devices.[4, 37]

## Experimental Section

The growth of $WSe_2$ thin film and in situ ARPES measurements were performed at the HERS endstation of beamline 10.0.1, Advanced Light Source, Lawrence Berkeley National Laboratory. The base pressures for the MBE system and ARPES system were ~2 × 10$^{-10}$ Torr and ~4 × 10$^{-11}$ Torr, respectively. ARPES data were taken with a Scienta R4000 electron analyzer, at a temperature of 60 K. The photon energy was set at 70 eV, with an energy and an angular resolution of 15 meV and 0.1°, respectively. The photon polarization direction was set to be 78° out of the incidence plane for an evenly distributed even and odd state signal. The size of the beam spot on the sample was ~150 μm × 200 μm. To protect the $WSe_2$ films from contamination of air during its transfer to the UHV-STM system, an ~ 100 Å amorphous Se capping layer was deposited on the sample before moving it out of the UHV-MBE chamber. Further annealing at ~ 300 ºC for one hour in the UHV-STM system was enough to remove the Se capping layer and uncover the pristine $WSe_2$ surface. Optical absorbance measurements were taken with a reflection configuration of a confocal microscope setup, using super-continuum white laser as the light source (focus spot ~ 2 μm).

## Supporting Information

Contribution of BLG band from the $WSe_2$ film in ARPES spectra, multiple Gaussian peaks fitting of the EDCs in ARPES spectra, the detailed surface doping effect in $MX_2$ films, optical absorption spectrum at different temperature, density of defects in $WSe_2$ films. This material is available free of charge via the Internet at http://pubs.acs.org.

## Acknowledgements

This work is supported by the US DOE, Office of Basic Energy Science, under contract no. DE-AC02-05CH11231 for ALS activities (growth and photoemission) and within the sp2 program (STM instrumentation development and operation), as well as by the US DOE Early Career Award No. DE-SC0003949 (optical measurements) and National Science Foundation Award No. EFMA-1542741 (image analysis). The work at the Stanford Institute for Materials and Energy Sciences and Stanford University is supported by the US DOE, Office of Basic Energy Sciences, under contract no. DE-AC02-76SF00515. The work at Oxford University is supported from a DARPA MESO project (no. 187 N66001-11-1-4105). The work at Pusan National University is supported by Max Planck Korea/POSTECH Research Initiative of the National Research Foundation (NRF) funded by the Ministry of Science, ICT & Future Planning under Project No. NRF-2011-0031558.

# Supplementary Information for

# Electronic Structure, Surface Doping and Optical Response in Epitaxial WSe$_2$ thin films

Yi Zhang[*], Miguel M. Ugeda, Chenhao Jin, Su-Fei Shi, Aaron J. Bradley, Ana Martín-Recio, Hyejin Ryu, Jonghwan Kim, Shujie Tang, Yeongkwan Kim, Bo Zhou, Choongyu Hwang, Yulin Chen, Feng Wang, Michael F. Crommie, Zahid Hussain, Zhi-Xun Shen, Sung-Kwan Mo[*]

Email: zhangyi@nju.edu.cn; SKMo@lbl.gov

## A: Contribution of BLG band from the WSe$_2$ films in ARPES spectra

In the angle-resolved photoemission spectroscopic (ARPES) measurement, we observed the Dirac cone from bilayer graphene (BLG) outside the Brillouin zone (BZ) of WSe$_2$. Figure S1a shows the BZ of BLG and WSe$_2$. The K point of WSe$_2$ is located at 1.27 Å$^{-1}$ away from the Γ point, whereas the K point of BLG (K$_G$) is 1.70 Å$^{-1}$ away from the T point, being located outside of the WSe$_2$ BZ. Figure S1b is the ARPES spectra of 1 ML WSe$_2$ film. The linear-dispersion bands of the two Dirac cones in BLG were clearly observed. This band structure of BLG is the same to the previous ARPES report of epitaxial BLG on SiC(0001) surface.[1] In the ARPES spectra of the 2 ML WSe$_2$ film, we found that the Dirac cones of BLG are completely absent (Figure S1c). This disappearance of the BLG signal means that the very short escape length of the photoelectrons minimizes the ARPES signal from the BLG substrate for 2ML or thicker WSe$_2$ films. A similar phenomena was also reported in the epitaxial MoSe$_2$ on BLG.[2] Since the BLG bands are located outside of and far away from the WSe$_2$ BZ in the same momentum direction (Figure S1b), it will not affect our observation and study of the WSe$_2$ valence band in the low energy scale along the Γ-K direction inside of the WSe$_2$ BZ.

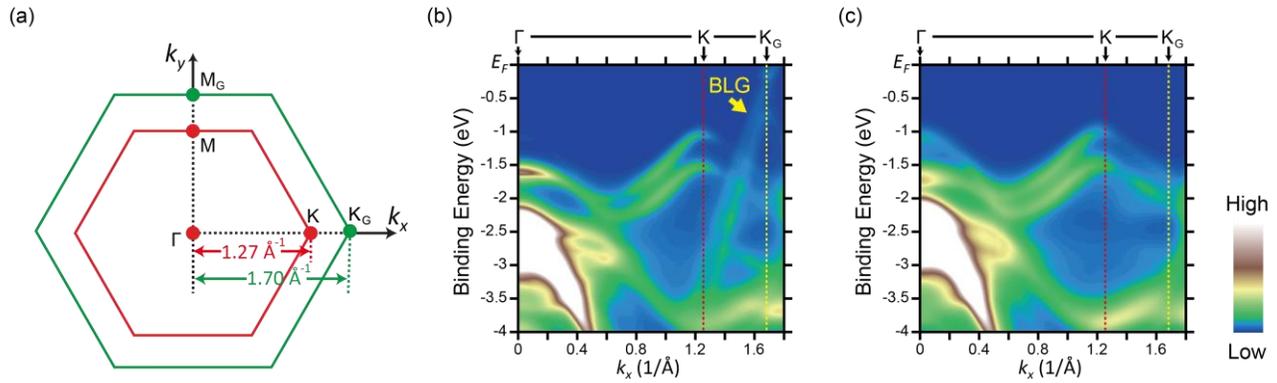

**Fig. S1 BLG band in ARPES spectra. (a)** Surface BZ of BLG (green hexagon) and WSe$_2$ (red hexagon). The K point of BLG BZ (K$_G$) is outside of the WSe$_2$ BZ. **(b) & (c)** ARPES spectra of (b) 1 ML WSe$_2$ and (c) 2 ML WSe$_2$ films. The red and yellow dotted lines indicate the K point of WSe$_2$ and K$_G$ point of BLG, respectively.

*B: Multiple Gaussian peaks fitting of the EDCs in ARPES spectra*

In order to precisely decide the energy position of the top valence band, multiple Gaussian peaks plus linear background was used to fit the energy distribution curves (EDCs) for each ARPES spectra of WSe$_2$ films. The colored dotted curves in Figure S2 show the fitting results of the EDCs at the Γ and K point. The fitting curves (cyan dotted curves) are in good agreement with the original EDCs (black curves). Each single Gaussian peak is plotted in different color (red, green and blue dotted curves), and the corresponding energy position of each peak is labeled by the same colored text. For the top valence band at the Γ point, the 1 ML WSe$_2$ has one peak at -1.629 eV (Figure S2a), the 2 ML WSe$_2$ has two peaks (Figure S2b), and the 3ML WSe$_2$ has three peaks (Figure S2c). For the 8 ML WSe$_2$, it is difficult to fit the Γ point EDC by using 8 Gaussian peaks curve. So we roughly determined the top of the valence band by cutting the half maximum of the EDC edge (red line in Figure S2d). In the K point EDC of 1 ML WSe$_2$, the two-Gaussian-peaks fitting curve is in good agreement with the EDC, indicating a well-defined ~475 meV spin-splitting of the top valence band at the K point (Figure S2e).

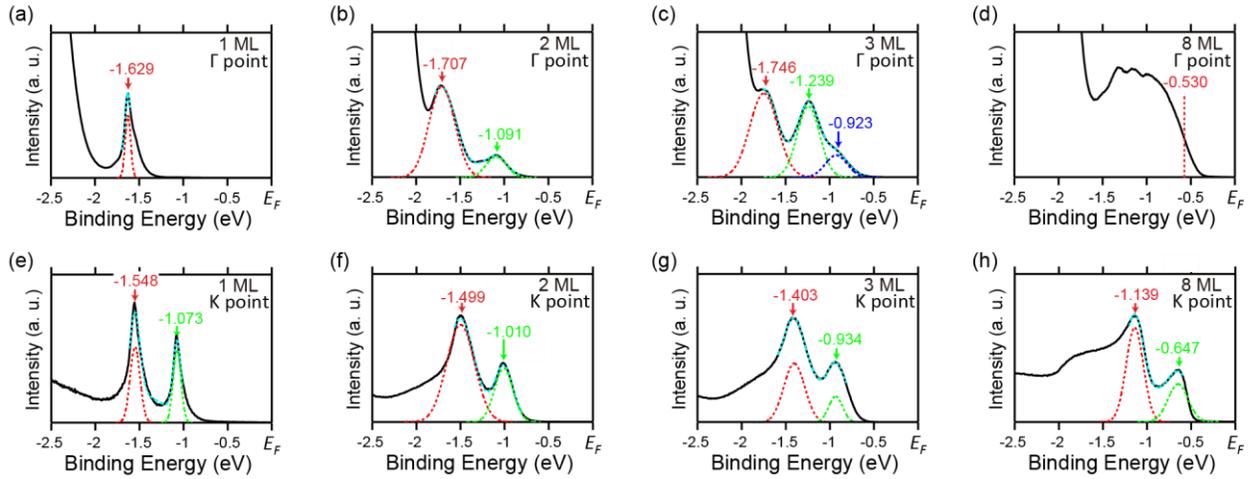

**Fig. S2 Multiple Gaussian peaks fitting for the EDCs. (a)-(d)** EDCs at the Γ point of the **(a)** 1 ML, **(b)** 2 ML, **(c)** 3 ML and **(d)** 8 ML WSe$_2$ films. **(e)-(h)** EDCs at the K point of the **(e)** 1 ML, **(f)** 2 ML, **(g)** 3 ML and **(h)** 8 ML WSe$_2$ films. The cyan dotted curves are the multiple Gaussian peaks fitting of the EDC peaks. The red, green and blue dotted curves are each single Gaussian peaks from the fitting curves. The energy position of each peak is labeled by the corresponding colored number. All the labeled numbers have unit of eV.

## C: The detailed surface doping effect in MX$_2$ films

We have introduced Na and K on the surface of WSe$_2$ films using a SAES Getters alkali metal dispenser. The amount of Na and K was controlled by the current applied to the dispenser ($I_d$) and the deposition time ($t_d$). During the surface doping, sample was keep at ~60 K.

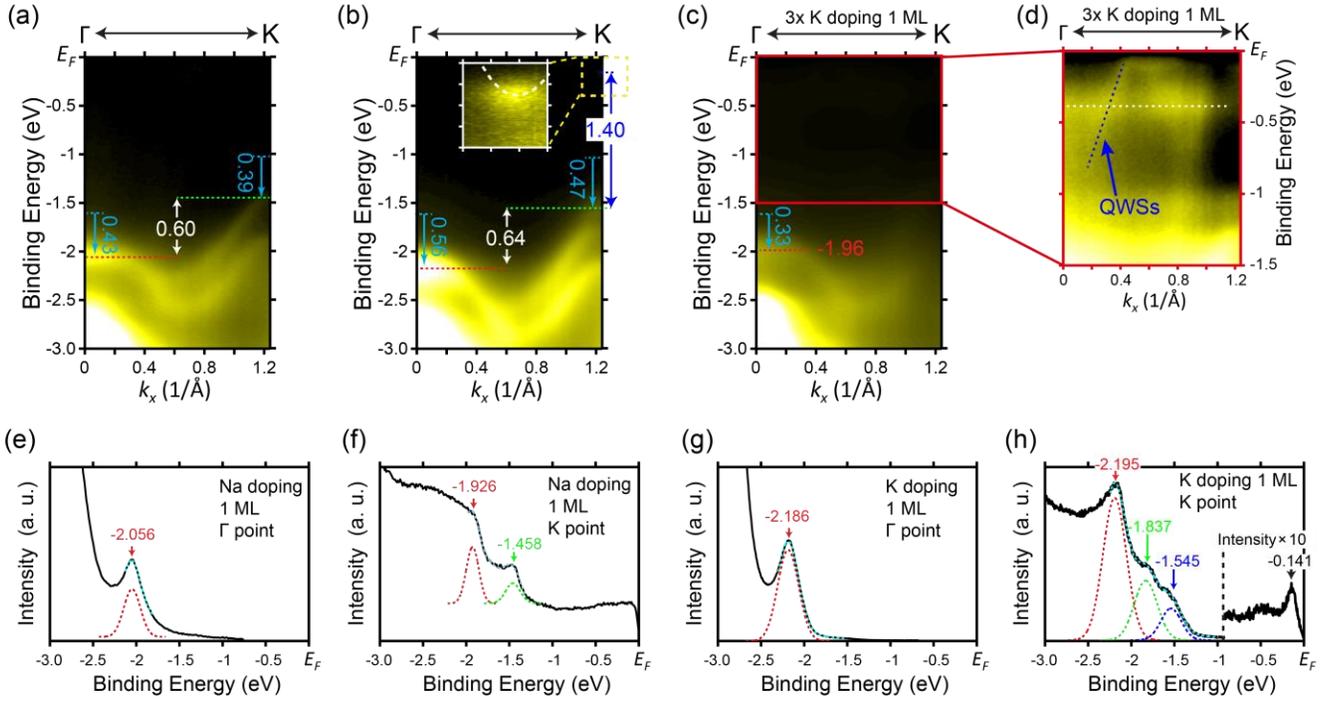

**Fig. S3 Surface doping effect of WSe$_2$ films. (a)-(c)** ARPES spectra of monolayer WSe$_2$ films with **(a)** Na surface doping ($I_d$ = 6.4 A, $t_d$ = 10 minutes), **(b)** K surface doping ($I_d$ = 6.0 A, $t_d$ = 10 minutes), and **(c)** 3 times K surface doping ($I_d$ = 6.0 A, $t_d$ = 30 minutes). Inset in **(b)** is the zoom-in spectra with 10 times enhanced intensity to make the CBM visible. The red and green dotted lines indicate the energy positions of the top valence band at the Γ and K point, respectively. Cyan arrows and numbers show the band movements at the Γ and K point after surface doping comparing to undoped films. **(d)** Zoom-in ARPES spectra of spectra **(c)** with 10 times enhanced intensity. The blue dotted line indicates the QWSs from monolayer K, and the white dotted line indicates the flat band from K cluster. **(e) & (f)** EDCs at the Γ point **(e)** and the K point **(f)** of Na surface doping ($I_d$ = 6.4 A, $t_d$ = 10 minutes) WSe$_2$ film, respectively. **(g) & (h)** EDCs at the Γ point **(g)** and the K point **(h)** of K surface doping ($I_d$ = 6.0 A, $t_d$ = 10 minutes) WSe$_2$ film, respectively. The cyan dotted curves are the multiple Gaussian peaks fitting of the EDC peaks. The red, green and blue dotted curves are each single Gaussian peaks from the fitting curves. The energy position of each peak is labeled by the corresponding colored number. All the labeled numbers have unit of eV.

Figure S3a is the ARPES spectra of monolayer WSe$_2$ with Na-doping ($I_d$ = 6.4 A, $t_d$ = 10 minutes). After the Na-doping, we found that the Fermi level moves upwards, i.e., the bands move downwards relative to the Fermi level. However, we noticed that the amount of valence band movement is different at the Γ and at the K point. At the Γ point, the top of the valence band moves 0.43 eV downwards after doping, while at the K point it only moves 0.39 eV. This enlarges the energy position difference between the Γ and K point ($E_K$-$E_Γ$ = 0.60 eV), and also indicates that the valence band is distorted after surface doping. A similar band distortion was also observed in the surface doping of the $MoS_2$[3] and $MoSe_2$ films (Figure S4). Such a momentum dependent band structure change could be due to a combination of the electronic potential originated from the surface change, the resulting band bending, and the varying responses against the potential from the bands with different orbital characters as well as different degrees of localization.[4] Despite the significant change in energy relative to the Fermi level, the movement due to the Na doping is still not large enough to make the CBM become visible, even with an increased amount of the doping.

To make the CBM become visible, surface doping of Potassium was found to be more effective. Figure S3b is the ARPES spectra of K-doped ($I_d$ = 6.0 A, $t_d$ = 10 minutes) monolayer $WSe_2$ film, pushing the top of the valence band 0.56 eV downwards at the Γ point, and 0.47 eV downwards at the K point. The heavier doping effect of K also brought larger changes in the electronic structure than Na-doped and undoped films. Figure S3c is the ARPES spectra of monolayer $WSe_2$ with 3 times more K surface doping ($I_d$ = 6.0 A, $t_d$ = 30 minutes). We found that the valence band at the Γ point moves back upwards in energy. When enhancing the intensity by 10 times in Figure S3d, we found additional bands in the $WSe_2$ band gap. One band indicated by blue dotted line is suggested to be the quantum well states (QWSs) due to the formation of a monolayer K film on the $WSe_2$ surface. Similar QWSs were also previously observed in K surface doping $MoS_2$ films.[5] Another band indicated by white dotted line is flat without momentum dispersion, suggested to be the impurity band as a result of the K clusters formation on the surface. The growth of a monolayer K film and K clusters weakens the electronic doping effect, and thus makes less bands movements, leaving the CBM still invisible in Figure S3c.

We used the same multiple Gaussian peaks fitting method on the EDCs of surface doping $WSe_2$ films to determine the energy positions of top valence bands and the energy movements after surface doping. Figures S3e-S3h show the EDCs and the fitting results of the Na surface doping ($I_d$ = 6.4 A, $t_d$ = 10 minutes) and K surface doping ($I_d$ = 6.0 A, $t_d$ = 10 minutes) monolayer $WSe_2$ films. In Figure S3h, by increasing the intensity by 10 times near the Fermi level, the peak of conduction band minimum (CBM) becomes visible at -0.141 eV at the K point. Thus we observed a direct band gap of 1.404 eV for K doped monolayer-$WSe_2$/BLG. Besides that, we also found that the valence band splitting at the K point has one more broaden shoulder in the EDC of K doping monolayer $WSe_2$, and the three-Gaussian-peaks fitting result is also in good agreement with the EDC line in Figure S3h. This suggests one more branch owing to the electric field induced Zeeman splitting from the K surface doping.[6]

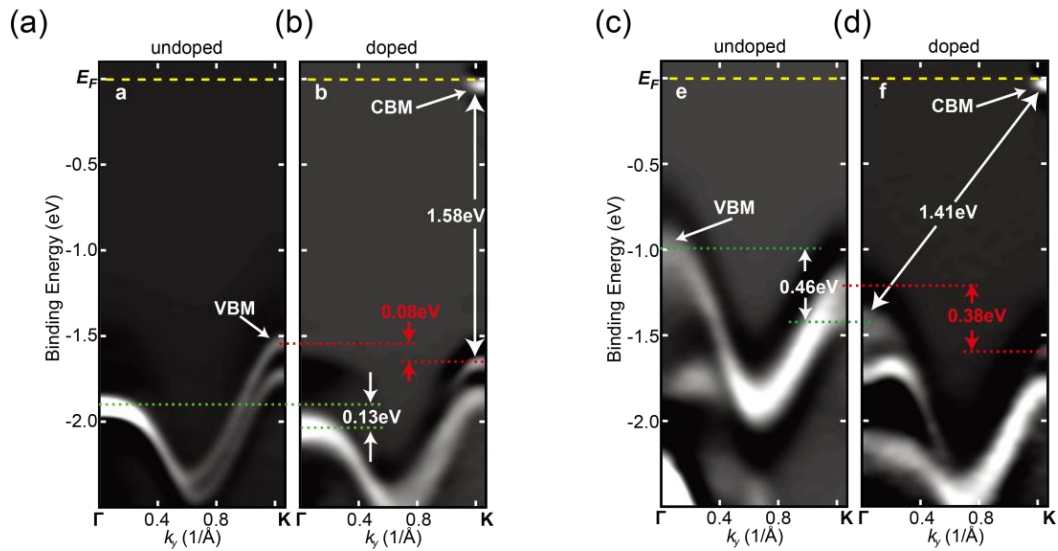

**Figure S4 Band distortion of K doped MoSe₂ films (a) & (b)** Second-derivative spectra of (a) undoped and (b) K surface doped monolayer MoSe₂, respectively. **(c) & (d)** Second-derivative spectra of (c) undoped and (d) K surface doped 8 ML MoSe₂. The green dashed lines indicate the band shift at the Γ point after surface doping. The red dashed lines indicate the band shift at the K point after surface doping.

In our previous work, we also did K surface doping on MoSe$_2$.[2] This surface doping also causes the band distortion like it does on WSe$_2$. But the amount of distortion is smaller than that on WSe$_2$ films. Figure S4 shows the band shift of MoSe$_2$ after K surface doping. Figures S4a and b are the ARPES spectra of monolayer MoSe$_2$ before and after doping, respectively. The valence band at the Γ point shifts downwards about 0.13 eV, but at the K point the amount of shifts is about 0.08 eV. Figures S4c and d are the ARPES spectra of 8 ML MoSe$_2$ before and after doping, respectively. The valence band shifts 0.46 eV at the Γ point but 0.38 eV at the K point. The different valence band shifts amount between the Γ point and K point indicates a similar band distortion caused by K surface doping in MoSe$_2$. Comparing to the surface doping on WSe$_2$, the K doped MoSe$_2$ films show smaller band distortion amount. Since the overall band shift in K-doped MoSe$_2$ films is also smaller than that in surface doped WSe$_2$, we believe that the dosing of dopants in MoSe$_2$ is less than that in WSe$_2$, and thus causes smaller distortion effect.

*D: Optical absorption spectrum at different temperature*

In Figure S5 we show the absorption spectrum taken on the 1 ML MoSe2 on BLG/SiC. The spectrum at 77 K (blue trace) clearly shows the resonances at 1.74 ± 0.01 eV and 2.17 eV ± 0.01 eV, which correspond to A and B exciton peaks. The room temperature spectrum (red trace) shows a shifted A exciton peak at 1.67 ± 0.01 eV. The absorption peak of B exciton is significantly broadened at room temperature and is hard to determine the peak position accurately. This red shift of exciton resonance energy at higher temperature has been observed previously and is likely due to temperature induced strain

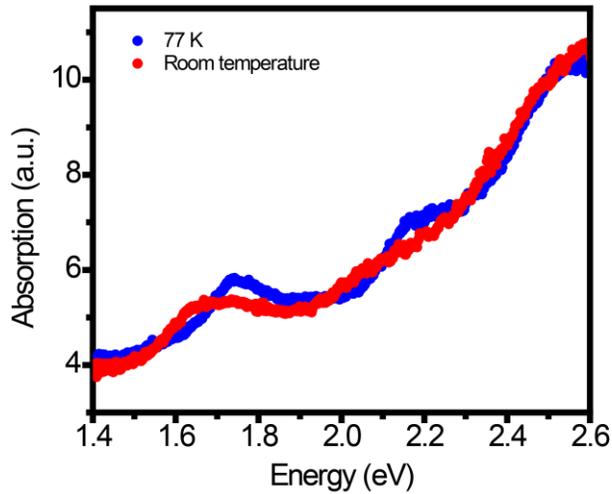

**Fig. S5 Optical absorption spectrum of monolayer WSe$_2$.** The blue trace is the spectrum taken at 77K, and the red trace is the spectrum taken at room temperature

change in the film.

*E: Density of defects in WSe$_2$ films*

In order to quantify the quality of our MBE grown WSe$_2$ films, we have estimated the density of point defects at the WSe$_2$ surface by analyzing our atomically resolved STM images. For this study we have used 31 STM images (T = 5 K) acquired at different bias voltages and set point currents. We have considered as point defect any protuberance or dip in the STM images as those shown in Figure S6a, regardless of their atomic nature. Figure S6b shows the number of point defects found in each of the 31 STM images as a function of the scanned surface area. A linear regression fit (red line) of the scattered values yields an estimation of the density of point defects ($\rho$) in our samples, which is found to be $\rho = (2.8 \pm 0.3) \cdot 10^{12}$ cm$^{-2}$. This density of defects estimated in our WSe$_2$ monolayers is almost three orders of magnitude smaller than the atomic surface density ($1.1 \cdot 10^{15}$ cm$^{-2}$), which proves the high quality of the presented MBE growth method. This estimation should be an upper bound value because the counting includes extrinsic surface defects such as adsorbates.

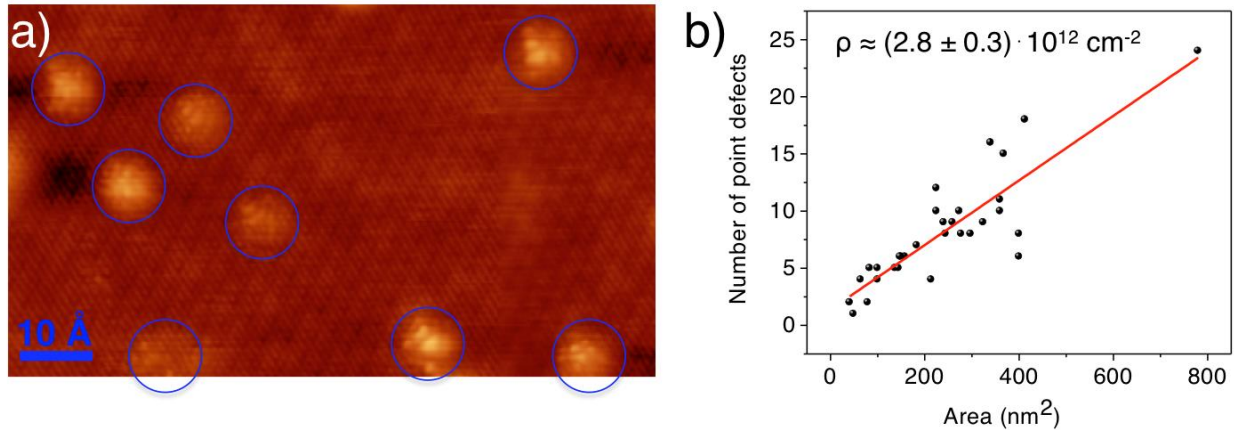

**Figure S6 Density of point defects on the surface of MBE grown WSe$_2$ monolayers** (a) STM image with point defects (blue circles) -(Vs = - 0.7 V, It = 500 pA, T = 5 K). (b) Number of point defects found in each STM image used in the analysis as a function of the scanned area.